\documentclass[12pt,onecolumn,oneside,a4paper,peerreview]{IEEEtran_kit}
\usepackage{cite}
\usepackage{float}
\usepackage{graphicx}
\usepackage{amsmath}
\usepackage{times}
\usepackage{graphicx}
\usepackage{latexsym}
\usepackage{graphicx}
\usepackage{bm}
\usepackage{amssymb}
\usepackage[center]{caption2}
\usepackage{stfloats}
\usepackage{cases}
\usepackage{array}
\usepackage{setspace}
\usepackage{fancyhdr}
\usepackage{color}
\usepackage{subfigure}
\usepackage{multirow}
\usepackage{amsmath}
\usepackage{cases}

\renewcommand{\QED}{\QEDopen}

\newtheorem{remark}{Remark}
\renewcommand{\baselinestretch}{0.97}

\newcaptionstyle{mystyle2}{%
\captionlabel $.$ \, \captiontext \par} \captionstyle{mystyle2}

\begin{document}
\markboth{IEEE Wireless Communications, vol. XX, no. XX, Month 2018.} {Zhong, Hu, Chen, Ng, \& Chen:
Spatial Modulation Assisted Multi-Antenna Non-Orthogonal Multiple Access \ldots}

\title{Spatial Modulation Assisted Multi-Antenna Non-Orthogonal Multiple Access}
\author{\normalsize
Caijun Zhong, \IEEEmembership{Senior Member, IEEE}, Xiaoling Hu, \IEEEmembership{Student Member, IEEE},\\ Xiaoming Chen, \IEEEmembership{Senior Member, IEEE}, Derrick Wing Kwan Ng, \IEEEmembership{Member, IEEE},\\ and
Zhaoyang Zhang, \IEEEmembership{Member, IEEE}
\thanks{Manuscript received March 11, 2017; revised August 09, 2017; accepted December 30, 2017. This work was supported by the National Natural Science Foundation of China under Grant 61725104, 61671406 and U1709219, the open research fund of National Mobile Communications Research Laboratory, Southeast University (No. 2018D07), the Zhejiang Provincial Natural Science Foundation of China under Grant LR15F010001, the National Science and Technology Major Project of China under Grant 2017ZX03001002-003 and 2018ZX03001017-002, and the Australian Research Council¡¯s Discovery Early
Career Researcher Award funding scheme under Grant DE170100137. \emph{(Corresponding author: Xiaoming Chen)}.}
\thanks{Caijun Zhong ({\tt caijunzhong@zju.edu.cn}), Xiaoling Hu, and Zhaoyang Zhang
are with the College of Information Science and Electronic Engineering,
Zhejiang University, Hangzhou, China. Xiaoming~Chen ({\tt chen\_xiaoming@zju.edu.cn}) is with the College of Information
Science and Electronic Engineering, Zhejiang University, Hangzhou,
China, and is also with the National Mobile Communications Research Laboratory,
Southeast University, Nanjing, China. Derrick Wing Kwan Ng ({\tt w.k.ng@unsw.edu.au}) is with the
School of Electrical Engineering and Telecommunications,
the University of New South Wales, NSW, Australia.}}\maketitle

\begin{abstract}
Multi-antenna non-orthogonal multiple
access (NOMA) is a promising technique to significantly improve the spectral efficiency and support massive access, which has received considerable interests
from academic and industry. This article first briefly introduces the basic idea of conventional multi-antenna NOMA technique, and then discusses the key limitations, namely, the high complexity of successive interference cancellation
(SIC) and the lack of fairness between the user with a strong channel gain
and the user with a weak channel gain. To address these problems, this
article proposes a novel spatial modulation (SM) assisted multi-antenna NOMA
technique, which avoids the use of SIC and is able to completely
cancel intra-cluster interference. Furthermore, simulation results are provided to validate the effectiveness of the proposed novel
technique compared to the conventional multi-antenna NOMA. Finally, this article points
out the key challenges and sheds light on the future research
directions of the SM assisted multi-antenna NOMA technique.
\end{abstract}
%
%\begin{keywords}{\centering}
%5G, multiple-antenna, NOMA, spatial modulation.
%\end{keywords}

\section{Introduction}\label{section:introduction}
To meet the exponential growth of mobile data traffic, future mobile networks are expected to deliver a $1,000$-fold capacity increase compared to the current wireless networks.
To make the full use of the current spectrum resource, it is imperative to develop spectral efficient technologies. In this context, non-orthogonal multiple access (NOMA), with huge potential of significantly boosting up the spectral efficiency, has received considerable attentions from both academia and industry recently
{\cite{NOMA1,NOMA2}}. For instance, multiuser superposition transmission (MUST), a downlink version of NOMA, has been proposed for the 3rd generation partnership project long-term evolution advanced (3GPP-LTE-A) networks. Furthermore, NOMA is also widely recognized as a key enabling technique for the fifth generation (5G) mobile systems.

%There are two different multiple access techniques, namely,
%orthogonal multiple access (OMA) and non-orthogonal multiple access
%(NOMA). By applying the orthogonal scheme such as time division
%multiple acesss (TDMA), orthogonal frequency-division multiple
%access (OFDMA) and code division access (CDMA), signals from
%different users are orthogonal to each other. OMA schemes eliminate
%mutual interference among the allocated users and allow relatively
%simple transceiver implementations. However, low-complexity
%implementation comes at the cost of efficiency.

In contrast to the conventional orthogonal multiple access
(OMA) such as time division multiple access (TDMA), frequency
division multiple access (FDMA), and code division multiple access
(CDMA), where each user occupies a distinct time/frequency/code channel, NOMA allows multiple users to share the same
time/frequency/code channel through superposition coding. Thus, NOMA
has the potential to substantially improve the spectral efficiency as well as support massive
access, which are two key requirements of 5G. On the
other hand, the use of superposition coding results in severe multiuser interference at the receivers. To mitigate the inter-user interference, successive interference cancellation (SIC) is usually performed at the receiver by
exploiting the signal strength gap between different users in the power domain \cite{SIC1}. However, the implementation of SIC incurs a high computational complexity. For instance, the user with the strongest channel gain needs to cancel all
the other users' interference, hence, if the number of users is large, the complexity of SIC is prohibitive. In addition, interference cancellation could be imperfect in practice, which results in error propagation and performance
degradation. Responding to this, user clustering has been proposed for the NOMA systems, where several users are grouped into a cluster, within which SIC is performed \cite{Clustering1}. To deal with the inter-cluster interference, multi-antenna NOMA appears to be a promising solution\cite{MIMONOMA3}. In particular, by exploiting the available channel state information at the transmitter (CSIT), spatial beamforming can be applied to effectively mitigate the inter-cluster interference.

While multi-antenna NOMA has significantly reduced the complexity at the user end, the use of SIC within each cluster still poses a significant challenge for many practical wireless systems, such as internet of things (IoT), where massive number of low-cost wireless devices with limited computational power is not able to perform sophisticated  operations. In addition, the use of SIC implies strong intra-cluster interference for the users with weak channel gains, hence, it is
difficult to guarantee user fairness. In an effort to circumvent the above issues in conventional multi-antenna NOMA systems, this article proposes a new spatial modulation (SM) assisted multi-antenna NOMA technique. For more details about the conventional SM, please refer to \cite{CONSM1,SM1}. Some previous researches have applied the SM techniques in the NOMA systems\cite{SMNOMA1,SMNOMA2}. However, these works are the combination of SM and the conventional NOMA. Thus, the challenges faced by the conventional NOMA, such as SIC and user fairness, also exist. In contrast, the proposed SM assisted multi-antenna NOMA technique is designed based on a novel SM technique, which avoids the use of SIC and improves user fairness.\footnote{In another independent patent issued recently, a similar idea was proposed in the context of WiFi \cite{Pat}. However, unlike \cite{Pat}, which merely deals with the dual user scenario, this article takes a step further by addressing the general multi-user scenario.}
Specifically, the proposed technique partitions the users into different pairs, and each pair shares a common channel by using a novel SM technique, which only requires a simple receiver and is able to completely avoid intra-cluster interference. The advantages of the proposed SM assisted multi-antenna NOMA technique are two-fold:
\begin{enumerate}
\item It avoids the use of SIC and thus has much lower computational complexity compared to the conventional multi-antenna NOMA.
\item It completely removes the intra-cluster interference, hence improves the performance and guarantees user fairness.
\end{enumerate}
%{\color{blue} We specially point out that the proposed SM assisted multi-antenna NOMA technique is not the simple combination of SM and NOMA, but a novel technique. Our proposed NOMA is achieved by a novel SM technique, while the conventional NOMA is achieved by power domain. Hence, it has essential difference from the conventional NOMA.}

The rest of this article is organized as follows. In Section
\ref{Conventional MIMO NOMA}, the basic concepts of the conventional
multi-antenna NOMA technique are presented. In Section \ref{spatial
modulation assisted MIMO NOMA}, we introduce in detail the SM
assisted multi-antenna NOMA technique, and present simulation results to illustrate its performance. In Section \ref{challenges and future direction},
we point out the key challenges and important future directions. Finally, we conclude this article in Section V.

\section{Conventional Multi-Antenna NOMA}\label{Conventional MIMO NOMA}
Combining multi-antenna and NOMA is an effective mean to satisfy the stringent requirements of high
spectral efficiency and massive access. As such, many research efforts have been made in designing various multi-antenna NOMA techniques. Among which, a common multi-antenna NOMA technique is that users in a cluster share the same spatial beam, and SIC is only
performed within the cluster \cite{MIMONOMA4}. Besides, spatial beams and transmit
powers are jointly optimized to mitigate both inter-cluster and intra-cluster interference. To better understand the advantages and
problems, we provide a brief introduction of conventional multi-antenna NOMA in this section.

Consider a downlink communication scenario, where a base station
(BS) equipped with $N_t$ antennas communicates with $K$
single-antenna users as illustrated in Fig.
\ref{fig:system_model_of_the_conventional_MIMO_NOMA}. In general, the multi-antenna NOMA downlink communication consists of
four key steps, namely, channel state information (CSI) acquisition,
user clustering, superposition coding, and SIC. In what follows, we introduce these key steps in
detail.

\subsection{CSI Acquisition}
CSI acquisition at the BS is necessary for the design of efficient multi-antenna NOMA downlink communication systems. Depending on the duplexing mode, the BS usually obtains the CSI of downlink channels in two ways. Specifically, in frequency duplex
division (FDD) systems, the BS transmits the pilot signal, and the users estimate the downlink channels, which are then reported to the BS through certain feedback channel. In time duplex division (TDD) systems, leveraging on the channel reciprocity, the BS acquires the
CSI by estimating the uplink channels. It is worth noticing that, due to channel estimation error and constrained feedback capacity, the BS may only have access to partial CSI.

\subsection{User Clustering}
In traditional power-domain NOMA systems, each receiver performs SIC
to remove the inter-user interference in order to improve the quality of the
received signal. However, if the number of users $K$ is large, the computational
complexity of SIC quickly becomes unbearable, and the residual interference after
SIC can be still very strong for users with weak channels. To tackle this issue, user clustering, where users are grouped into distinct clusters, is proposed, and SIC is only carried out within each cluster. User clustering not only decreases the complexity of SIC, but also alleviates the residual intra-cluster
interference due to reduced number of users within a cluster. However, the above advantages come at the price of introducing inter-cluster interference. Therefore, the key of user clustering is to achieve a fine balance
between intra-cluster interference and inter-cluster interference by
appropriately adjusting the number of clusters and the number of
users in each cluster.

\subsection{Superposition Coding}
In layman's term, superposition coding is equivalent to a weighted
sum of the user signals with transmit powers being the
weights. However, superposition coding is more complicated in multi-antenna NOMA systems due to the requirement of joint design of the spatial beamforming vectors and transmit powers. Specifically, transmit beamforming is used to suppress the inter-cluster interference. With sufficient number of BS antennas and full CSI, simple zero-forcing beamforming can completely remove the inter-cluster interference. Afterwards, power allocation is used to further
decrease intra-cluster interference. However, since the users in a cluster share the same spatial beam, the effective channel gains of users may be affected. Thus, a joint design of spatial beams and transmit powers is of critical importance to improve the performance of multi-antenna NOMA systems.

\subsection{Successive Interference Cancellation}
SIC is a key technique of traditional power-domain NOMA system for
performance improvement. In particular, after receiving the
superposition coded signal, each user first decodes the signals of
the users with weaker channel gain in the same cluster, then removes
the interference from these weak users before demodulating its own
signal. Thus, the user with the strongest channel gain is required to cancel
all intra-cluster interference leading to a high SIC complexity. While
the user with the weakest channel gain has a low implementation
complexity, but receives interference from all the other users in
the same cluster. Therefore, there are significant difference between the achievable rates of different users.

\section{Spatial Modulation Assisted Multi-Antenna NOMA}\label{spatial modulation assisted MIMO NOMA}
As discussed above, conventional multi-antenna NONA systems mainly exploit superposition coding to realize channel sharing, and adopt
SIC to mitigate intra-cluster interference. However, the implementation of superposition coding and SIC has high computational complexity. Thus, it is of paramount importance to devise a low-complexity multi-antenna NOMA technique. Motivated by this, we propose a novel spatial modulation (SM) assisted multi-antenna NOMA technique in this section. We first start with a simple two-user multi-antenna NOMA technique, then extend it to a general multiuser case and discuss the associated key techniques. Finally, we show the performance of the proposed SM assisted multi-antenna NOMA technique by numerical simulations.

\subsection{Two-User SM Assisted Multi-Antenna NOMA}\label{Two-User SM Assisted MIMO NOMA}
Prior to the design of SM assisted multi-antenna NOMA technique, we first
provide a brief introduction of the basic concept of conventional
SM.

A distinct characteristic of SM is that it allows additional
information being conveyed through antenna selection, and thus the
overall spectral efficiency can be improved \cite{SM1}, \cite{SM2}.
Specifically, SM maps the information bits into two information
carrying units: (1) a symbol that is chosen from a constellation
diagram and (2) the index of the selected antenna for transmission.
To clarify the conventional SM, let us consider a simple
point-to-point SM communication system as depicted in Fig.
\ref{fig:system_model_of_conventional_spatial_modulation}, where a
BS equipped with $N_t$ antennas sends amplitude and phase modulation (APM)
symbols to a user equipment (UE) with $N_r$ antennas, and a QAM constellation
diagram of size $M$ is adopted for signal modulation. The
information bit stream vector $\bf{b}$ is split into two sub-vectors denoted by
${\bf{b}}_{1}$ and ${\bf{b}}_{2}$, respectively. The sub-vector
${\bf{b}}_{1}$ with $\operatorname{log}_{2}\left(N_t\right)$ bits is
used to determine the index $i$ of the activated antenna for
transmission, while ${\bf{b}}_{2}$ with $\operatorname{log}_{2}
\left(M\right)$ bits is mapped into an APM symbol $x$ chosen from a
$M$-QAM constellation diagram. Then, the symbol $x$ is emitted from
the activated antenna $i$. Finally, the UE adopts maximum likelihood
(ML) detection scheme to recovery both the transmitted symbol and
the index of activated antenna.

Inspired by the above SM technique, we propose a novel two-user
SM assisted MIMO-NONA technique. The basic procedure of this
technique is shown in Fig.
\ref{fig:system_model_of_novel_spatial_modulation}, where the
information for one user is mapped to the index of transmit antenna,
while the information for the other user is mapped into an APM
symbol. At the receiver side, UE1 uses a simple maximum ratio
combining (MRC) scheme to estimate the index of the activated antenna. Specifically, the received signal is multiplied by the Hermititian conjugate of ${\bf{h}}_{i,1}, i=1,...,N_t$, where ${\bf{h}}_{i,1}$ denotes the channel between UE1 and the $i$-th antenna of the transmitter, yielding an $N_t$-dimensional vector $\bf{g}$. Then, the index of the activated antenna is the position of that element in ${\bf{g}}$ with the largest absolute value.
At UE2, the ML scheme is used to calculate the squared Euclidean distance between the received signal and the signal ${\bf{h}}_{i,2}x_n, i=1,...,N_t, n=1,...,M$, where ${\bf{h}}_{i,2}$ denotes the channel between UE2 and the $i$-th antenna of the transmitter, and $x_n$ is the $n$-th symbol of the $M$-QAM constellation diagram. Moreover, the combination of the index and the symbol, i.e., $\left(i,x_n\right)$, which results in the minimum squared Euclidean distance, is the estimated result.

By doing so, the information for
the two users can be conveyed in the same time/frequency/code
channel.

\begin{remark}
The advantages of the proposed SM assisted multi-antenna NOMA
scheme are two-fold. First, there is no intra-cluster interference
between the two users in the same cluster, thus it is possible to
improve the performance. Second, the complicated SIC signal processing at the receiver
is avoided, hence it  decreases the computational
complexity and reduces the decoding latency of multi-antenna NOMA systems.
\end{remark}

\subsection{Multiuser SM Assisted Multi-Antenna NOMA}\label{Multiuser SM Assisted MIMO NOMA}
In practical systems, there may be more than two users accessing the
same spectrum simultaneously. Thus, it is necessary to design a multiuser SM
assisted multi-antenna NOMA scheme. Now, we extend the previously proposed scheme to the case of multiple users.

Consider a single-cell multiuser downlink MIMO system, where the BS
equipped with $N_t$ antennas serves $K$ pairs of UEs on the same
time/frequency/code channel. It is assumed that each UE has $N_r$
antennas. Different from the traditional multiuser MIMO systems,
where all BS antennas serve all users simultaneously, in the
proposed multiuser SM multi-antenna NOMA system, the $N_t$ BS antennas are
divided into $K$ groups, each having ${N_t}/K$ antennas and serving
a specific UE pair. Without loss of generality, the $i$-th BS
antenna group serves the $i$-th UE pair denoted by
${\operatorname{UE}}_{i,1}$ and ${\operatorname{UE}}_{i,2}$,
respectively. The system model is shown in Fig. \ref{fig:system
model of the spatial modulation assisted MIMO NOMA}.

In each group, the signal to be transmitted is processed similar to
the two-user multi-antenna NOMA. However, due to the open nature of wireless
channels, there exists inter-group interference at UEs, resulting in
a performance loss. In order to improve the performance of multiuser
multi-antenna NOMA systems, it is necessary to carry out interference
mitigation technique at the UE, or the BS, or the both.
Since BS is equipped with multiple antennas, it is possible to make use of transmit beamforming to
tackle the inter-group interference. However, for the proposed multiuser SM assisted multi-antenna NOMA, part of the information is encoded according to the channel impulse responses (CIRs). Hence, the conventional interference mitigation techniques, such as zero forcing (ZF) or minimum mean square error (MMSE), are no longer applicable. Instead, new transmit beamforming schemes which can cancel the interference and preserve the information encoded as constellation symbols and transmit antenna indexes should be used \cite{MUSM1,MUSM2}. After interference mitigation, UE1 and UE2
in each group use ML detection to recovery the desired
information.

%{\color{blue}
%\begin{remark}
%It is worth noting that the proposed NOMA separates the users by a
%novel SM technique, but not in the power domain. Hence, the proposed
%NOMA is quite different from the conventional NOMA. As illustrated in
%Section III-A, for the proposed novel SM technique, the information
%bits for one user are represented by the index of transmit antenna,
%while the information bits for the other user are mapped into an APM
%symbol simultaneously. In this way, the two users can be served at
%the same time/frequency/code, for which we called this technique as
%two-user SM assisted multi-antenna NOMA. Then, by user pairing,
%antenna allocation, and spatial beamforming, two-user SM assisted
%multi-antenna NOMA is extended to multi-user SM assisted
%multi-antenna NOMA, where multiple users can be served at the same
%time/frequency/code.
%\end{remark}
%}

\subsection{Key Techniques of SM Assisted Multi-Antenna NOMA}
In the proposed multiuser SM assisted multi-antenna NOMA scheme, the users
are partitioned into $K$ user pairs, and each user pair performs SM
to share a time/frequency/code channel, while inter-pair
interference is controlled by some interference mitigation
techniques. Thus, the following techniques should be carefully
designed according to the characteristics of the multiuser SM
assisted multi-antenna NOMA scheme.

\subsubsection{User Pairing}
For the proposed multiuser SM assisted multi-antenna NOMA scheme, there is no intra-pair
interference between the two users, but there exists inter-pair
interference. Thus, a proper design of user pairing to mitigate
inter-pair interference is very important to improve the overall
performance. Since the inter-pair interference is suppressed by
spatial beamforming, it is desirable to pair the two users with
similar signal directions together, such that the effective channel
gains can be enhanced for the both users and to facilitate the cancellation of inter-pair
interference. In addition, inspired by the fact that detecting the APM symbol may result in high bit-error-rate (BER), especially at low signal-to-noise ratio (SNR), while the BER caused by detecting the index of the activated antenna is less influenced by the SNR\cite{SM1}, we propose that for the two paired users, the one with stronger channel is to detect the transmitted symbol, and another user is to detect the index of the activated antenna. In this way, for the user detecting the transmitted symbol, the BER is greatly improved due to its strong channel condition. Besides, for another user detecting the index of the activated antenna, there is only a small increase in BER, since it is less influenced by the SNR. Hence, the overall performance can be improved.
%{\color{blue}Besides, inspired by the fact that detecting the transmitted symbol may result in high bit error rate, especially at low SNR, while bit error rate caused by detecting the index of the activated antenna is less influenced by the SNR, we propose another user pairing scheme: the user with poor channel is paired with the user with strong channel. Meanwhile, the former detects the index of the activated antenna, and the latter detects the transmitted symbol. In this way, for the user detecting the transmitted symbol, the bit error rate is greatly reduced due to its strong channel. And, for another user detecting the index of the activated antenna, the bit error rate has small increase, since it is less influenced by the SNR.}

\subsubsection{Antenna Allocation}
For the proposed multiuser SM assisted multi-antenna NOMA scheme, different user pairs are effectively served by different sets of BS antennas. Since each antenna has a distinct channel gain towards each user pair, antenna selection can have a significant impact on the user performance. In addition, the number of antennas for a pair determines the information rate of UE1. Therefore, in order to optimize the overall performance, it
is necessary to allocate the BS antennas intelligently according to the channel conditions and user performance requirements. In general, the optimal antenna allocation problem is a combinatorial optimization problem, hence can be only solved by the exhaustive search method.
%However, in future 5G systems, the
%BS is usually equipped with a large-scale antenna array
%\cite{MassiveMIMO}, and hence the complexity of exhaustive search
%may be prohibitive. Therefore, it is necessary to design a
%low-complexity antenna allocation strategy.

\subsubsection{Inter-pair Interference Cancellation}
The inter-pair interference may substantially degrade the performance of the multiuser SM assisted multi-antenna NOMA
scheme, hence must be properly handled. Capitalizing on the multiple antennas at the BS, transmit beamforming appears a promising candidate for
interference mitigation. As mentioned above, the BS may only be able to obtain partial CSI, hence it is necessary to design robust interference cancelation
strategies for optimizing the performance in the worst case.

\subsection{Simulation Results}\label{simulation results}
We now present numerical simulation results to compare the performance of the SM assisted multi-antenna NOMA with the conventional multi-antenna NOMA.
% For all simulations, Rayleigh fading and additive white Gaussian noise (AWGN) channel are considered.
We consider the system with bandwidth of 4.32 MHz and
noise density of -169 dBm/Hz. For all simulations, the channel is
modeled as a product of path loss and Rayleigh fading with zero mean
and unit variance. The path loss model used is $128.1 +
37.6\operatorname{log}_{10} \left(r\right)$ dB, where $r$ (km) is
the distance between the transmitter and the receive. We further
assume that the distance between the BS and UE1/UE2 is 0.15 km and
0.1 km, respectively, where UE1 and UE2 are any two paired users.
Also, we assume that the SM assisted multi-antenna NOMA system has
finite alphabet inputs, while the conventional multi-antenna NOMA
uses Gaussian inputs. For simplicity, equal power allocation policy
is used for the SM assisted multi-antenna NOMA system. Finally, we
use $N_u$ to denote the number of users, $N_s$ to denote the size of
the QAM constellation diagram. All the simulation curves are
generated by averaging over $100,000$ independent channel
realizations.

We first compare the sum rate performance of the two multi-antenna NOMA
schemes with different $N_u$ and $N_s$. From Fig.
\ref{fig:simulation of sumrate}, we obtain the following key observations:

\begin{itemize}
\item  When $N_t=8$, $N_u=8$, and $N_s=64$, the SM
assisted multi-antenna NOMA system achieves a higher ergodic sum rate over
the conventional multi-antenna NOMA in the low and moderate SNR regime. When the number of users is
decreased to $4$, i.e., $N_u=4$, the ergodic sum rate of the SM assisted multi-antenna NOMA becomes smaller than that of the conventional multi-antenna NOMA. The reason is that, with small number of users, the intra-cluster user interference is less severe, hence, does not significantly degrade the sum rate performance. When the number of users increases, the intra-cluster user interference becomes a major performance limiting factor for the conventional multi-antenna NOMA system.

\item For a fixed $N_t$ and $N_s$, the ergodic sum
rate of the SM assisted multi-antenna NOMA gradually improves as $N_u$ increases. In other words, the SM assisted multi-antenna NOMA system achieves a superior performance for a large number of users, which makes the proposed scheme appealing since the number of
devices accessing to the communication network will grow rapidly in the future 5G mobile systems.

\item For fixed $N_t$ and $N_u$, the ergodic sum rate of the SM assisted multi-antenna NOMA depends heavily with the modulation level $N_s$. In the low SNR regime, increasing the modulation level is not able to contribute to the sum rate due to elevated decoding error. In contrast, in the high SNR regime, increasing the modulation level may significantly improves the sum rate. Hence, adaptive modulation scheme should be used according to the operating environments.

\item For fixed $N_u$ and $N_s$, the ergodic sum rate of the SM assisted multi-antenna NOMA improves as $N_t$ increases,
which indicates that it is always desirable to deploy more transmit antennas in terms of improving the sum rate performance.
\end{itemize}

Having investigated the sum rate performance, we now compare the ergodic rate of
the worst user in the SM assisted multi-antenna NOMA system with that in the
conventional multi-antenna NOMA system
in Fig. \ref{fig:simulation of worst rate}.

When $N_u=8$, the rate of the worst user in the SM assisted multi-antenna NOMA system is lower than that of the conventional multi-antenna NOMA system. In contrast, when $N_u=4$, the rate of the worst user in the SM assisted multi-antenna NOMA system is slightly inferior to that of the conventional multi-antenna NOMA system in the low SNR regime, while becomes superior in the high SNR regime. The reason is that, in the low SNR regime, the worst user in the conventional multi-antenna NOMA system obtains more spatial diversity gains, while in the high SNR regime, it is interference limited, hence, its achievable rate saturates quickly. In contrast, the rate of the worst user in the SM assisted multi-antenna NOMA system is constrained by the number of antennas for each pair in the high SNR regime. For instance, the highest rate is 2 bits/s/Hz when $N_u=4$ since each pair is served by 4 antennas, and 1 bits/s/Hz when $N_u=8$ since each pair is served by 2 antennas. Also, as the number of user increases, the rate of the worst users in both cases degrades. This is intuitive since increasing the number of users implies higher interference for the conventional NOMA system, and reduced number of antennas for each pair for the SM assisted NOMA system. Finally, we see that a higher modulation level reduces the rate of the worst user. This is because a higher modulation level may result in a larger probability of incorrect index detection at the worst user.

\section{Challenges and Future Direction}\label{challenges and future direction}
In the previous section, we have briefly introduced the basic idea and concept of SM assisted multi-antenna NOMA. However, to fully unleash the potential of the proposed method, there are still many practical challenges need to be tackled, as elaborated below.
\subsection{Joint User Pairing and Antenna Partition}\label{user clustering}
In the proposed multi-user SM assisted multi-antenna NOMA system, the two users in each pair adopt different modulation formats, i.e., QAM modulation and spatial modulation, hence, the user paring methods proposed for users with same type of modulation format may no longer be suitable, as such, it is important to design new user pairing method tuned for the considered scenario. In addition, the users and antennas are coupled through the propagation channels, hence, to achieve optimal performance, it is imperative to jointly optimize the user pairing and antenna partition.

\subsection{Adaptive Modulation}
As illustrated in Fig. \ref{fig:simulation of sumrate}, the achievable sum rate depends heavily on the operating SNRs and modulation level $N_s$. Hence, adaptive modulation should be applied to improve the sum rate performance. However, different from the conventional single user adaptive modulation scheme, where the change of modulation level does not affect other users, in the current system, the change of modulation level of UE2 may also affect the signal detection performance of UE1. Therefore, novel adaptive modulation scheme should be designed, which takes into account of the channel conditions and QoS requirements of both users.

\subsection{Joint Beamforming Design and Power Control}\label{power control}
Beamforming design and power control are two key methods to tackle the inter-pair interference for the multi-user SM assisted multi-antenna NOMA system. The primary goal of beamforming design is to suppress the inter-pair interference, nevertheless, it affects the effective channel gain of the users in each pair.  Therefore, beamforming design and power control should be jointly considered to achieve optimal performance.

\subsection{Imperfect CSI}\label{limited feedback}
In practice, acquiring the CSI at transmitter incurs significant system overhead, and due to practical constraints, it is likely that the CSI at the transmit is imperfect. Since the user pairing, antenna partition, beamforming design and power control all require the availability of CSI at the transmitter, the imperfection of the CSI will greatly degrades the performance of the system. Therefore, there is a pressing need to develop novel design theories and practical algorithms to deal with the case with imperfect CSI.

\subsection{Massive MIMO}\label{massive MIMO}
Since massive MIMO has the great potential to enhance the capacity
of communication systems, it is highly desirable to combine the novel SM assisted NOMA with massive MIMO. However, to reap the benefits of massive MIMO, a number of issues need to be addressed. For instance, when the number of transmit antennas is very large, channel estimation in frequency division duplexing (FDD) massive MIMO systems becomes challenging. Therefore, it is especially important to study the impact of imperfect CSI. As a low complexity alternative, limited feedback scheme is of particular interests \cite{X.Chen}. In addition, with
large-scale antenna array, optimal antenna partition and power control become much more difficult to deal with. Hence, it is also desirable to devise low-complexity suboptimal antenna partition and power control strategies.

\section{Conclusion}
This article proposed a novel SM assisted multi-antenna NOMA scheme for the
future 5G mobile communication systems. Different from the
conventional multi-antenna NOMA scheme, the proposed scheme makes use of SM
to realize channel sharing between two users. Furthermore, through
antenna partition, the SM assisted multi-antenna NOMA scheme was extended to
support multi-user access. The proposed SM assisted multi-antenna NOMA scheme
is capable of avoiding the use of SIC at the receiver, and thus
significantly decrease the computational complexity.\footnote{Although only the complexity of the receiver is mentioned in this paper, taking into account the complexity of other techniques, such as joint user pairing and antenna partition, joint beamforming design and power control, as well as adaptive modulation, the overall complexity of the proposed SM assisted multi-antenna NOMA systems is still lower than that of the conventional NOMA systems.} Moreover, it is able to completely cancel intra-cluster interference, which is
helpful to achieve user fairness.

\nocite{*}
\bibliographystyle{IEEE}
\begin{footnotesize}

\begin{biographynophoto}{Caijun Zhong}(S'07-M'10-SM'14) received the B.S. degree in Information Engineering from the Xi'an Jiaotong University, Xi'an, China, in 2004, and the M.S. degree in Information Security in 2006, Ph.D. degree in Telecommunications in 2010, both from University College London, London, United Kingdom. From September 2009 to September 2011, he was a research fellow at the Institute for Electronics, Communications and Information Technologies (ECIT), Queen¡¯s University Belfast, Belfast, UK. Since September 2011, he has been with Zhejiang University, Hangzhou, China, where he is currently an associate professor. His research interests include massive MIMO, wireless power transfer and backscatter communications. Dr. Zhong is an Editor of the \textsc{IEEE Transactions on Wireless Communications}, \textsc{IEEE Communications Letters}, \textsc{EURASIP Journal on Wireless Communications and Networking}, and \textsc{Journal of Communications and Networks}. He is the recipient of the 2013 IEEE ComSoc Asia-Pacific Outstanding Young Researcher Award. He and his coauthors received the Best Paper Award at the WCSP 2013 and SigTelCom 2017.
\end{biographynophoto}

\begin{biographynophoto}{Xiaoling Hu}
(S'15) received the B.S. degree in electronics and information engineering from the Dalian University of Technology, Dalian, China, in 2016. She is currently pursuing the Ph.D. degree in information and communication engineering with Zhejiang University. Her research interests include massive MIMO systems and NOMA communications.
\end{biographynophoto}

\begin{biographynophoto}{Xiaoming Chen}
(M'10-SM'14) received the B.Sc. degree from Hohai University in 2005, the M.Sc. degree from Nanjing University of Science and Technology in 2007 and the Ph. D. degree from Zhejiang University in 2011, all in
electronic engineering. Since November 2016, he has been with the College of Information Science and Electronic Engineering, Zhejiang University, Hangzhou, China, where he is now a Young Professor. From March 2011 to October 2016, he was with Nanjing University of Aeronautics and Astronautics, Nanjing, China. From February 2015 to June 2016, he was a Humboldt Research Fellow at the Institute for Digital Communications, Friedrich-Alexander-University Erlangen-N\"urnberg (FAU), Germany. His research interests mainly focus on massive access, non-orthogonal multiple access, physical layer security, and wireless power transfer. Dr. Chen is currently serving as an Associate Editor for the \textsc{IEEE Access} and an Editor for the \textsc{IEEE Communications Letters}. He was an Exemplary Reviewer for \textsc{IEEE Communications Letters} in 2014, and \textsc{IEEE Transactions on Communications} in 2015, and 2016.
\end{biographynophoto}

\begin{biographynophoto}{Derrick Wing Kwan Ng}(S'06-M'12)
received the bachelor degree with first class honors and the Master of Philosophy (M.Phil.) degree in electronic engineering from the Hong Kong University of Science and Technology (HKUST) in 2006 and 2008, respectively. He received his Ph.D. degree from the University of British Columbia (UBC) in 2012. In the summer of 2011 and spring of 2012, he was a visiting scholar at the Centre Tecnol\`{o}gic de Telecomunicacions de Catalunya - Hong Kong (CTTC-HK). He was a senior postdoctoral fellow at the Institute for Digital Communications, Friedrich-Alexander-University Erlangen-N\"urnberg (FAU), Germany. He is now working as a Senior Lecturer and an ARC DECRA Research Fellow at the University of New South Wales, Sydney, Australia.  His research interests include convex and non-convex optimization, physical layer security, wireless information and power transfer, and green (energy-efficient) wireless communications. Dr. Ng received the Best Paper Awards at the IEEE International Conference on Computing, Networking and Communications (ICNC) 2016,  IEEE Wireless Communications and Networking Conference (WCNC) 2012, the IEEE Global Telecommunication Conference (Globecom) 2011, and the IEEE Third International Conference on Communications and Networking in China 2008. He has served as an editorial assistant to the Editor-in-Chief of the IEEE Transactions on Communications since Jan. 2012.  He is currently an Editor of the IEEE Communications Letters, the IEEE Transactions on Wireless Communications, and the IEEE Transactions on Green Communications and Networking. He was honored as an Exemplary Reviewer of the IEEE Transactions on Communications in 2015, the top reviewer of IEEE Transactions on Vehicular Technology in 2014 and 2016, and an Exemplary Reviewer of the IEEE Wireless Communications Letters for 2012, 2014, and 2015.
\end{biographynophoto}

\begin{biographynophoto}{Zhaoyang Zhang}
(M'00) received his Ph.D. degree in communication and information systems from Zhejiang University, China, in 1998. He is currently a full professor with the College of Information Science and Electronic Engineering, Zhejiang University. His current research interests are mainly focused on some fundamental and interdisciplinary aspects of communication and compuation, such as communication-computation convergnce and network intelligence, theoretic and algorithmic foundations for Internet-of-Things (IoT) and Internet-of-Data (IoD), etc. He has co-authored more than 200 refereed international journal and conference papers as well as two books. He was awarded the National Natural Science Fund for Distinguished Young Scholars from NSFC in 2017, and was a co-recipient of five international conference best paper awards. He is currently serving as Editor for IEEE Transactions on Communications, IET Communications and some other international journals. He served as General Chair, TPC Co-Chair or Symposium Co-Chair for many international conferences like VTC-Spring 2017 HMWC Workshop, WCSP 2018/2013 and Globecom 2014 Wireless Communications Symposium, etc.
\end{biographynophoto}

\newpage

\begin{figure}[!ht]
  \centering
  \includegraphics[scale=0.45]{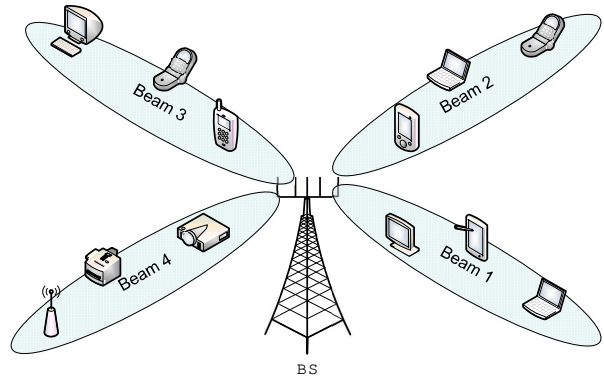}
%  %\hspace{1in}
%  \subfigure[]{\label{fig:1b}\includegraphics[scale=0.6]{system_model_version222.eps}}
  \caption{Schematic model of conventional multi-antenna NOMA.}
  \label{fig:system_model_of_the_conventional_MIMO_NOMA}
\end{figure}

\begin{figure}[!ht]
  \centering
  \includegraphics[scale=0.75]{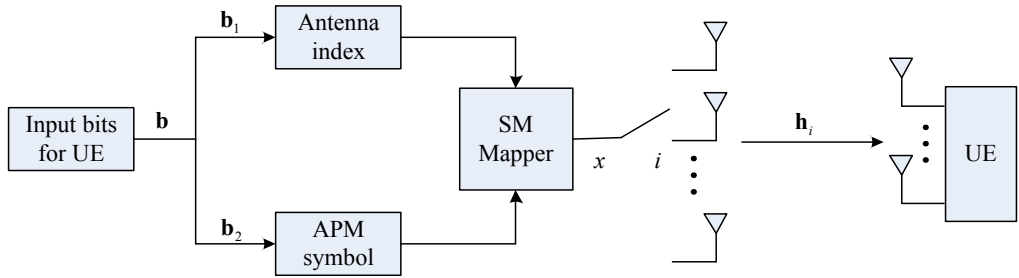}
%  %\hspace{1in}
%  \subfigure[]{\label{fig:1b}\includegraphics[scale=0.6]{system_model_version222.eps}}
  \caption{Schematic model of the conventional spatial modulation system.}
  \label{fig:system_model_of_conventional_spatial_modulation}
\end{figure}

\begin{figure}[!ht]
  \centering
  \includegraphics[scale=0.75]{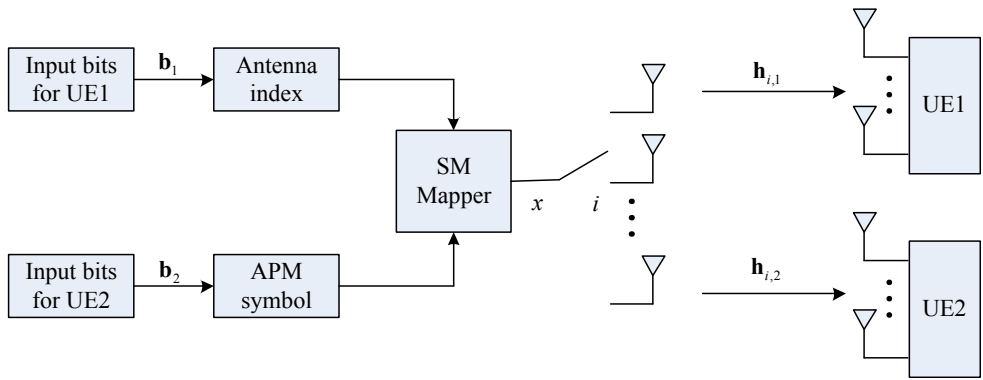}
%  %\hspace{1in}
%  \subfigure[]{\label{fig:1b}\includegraphics[scale=0.6]{system_model_version222.eps}}
  \caption{System model of a two-user SM assisted multi-antenna NOMA.}
  \label{fig:system_model_of_novel_spatial_modulation}
\end{figure}

\begin{figure}[!ht]
  \centering
  \includegraphics[scale=0.75]{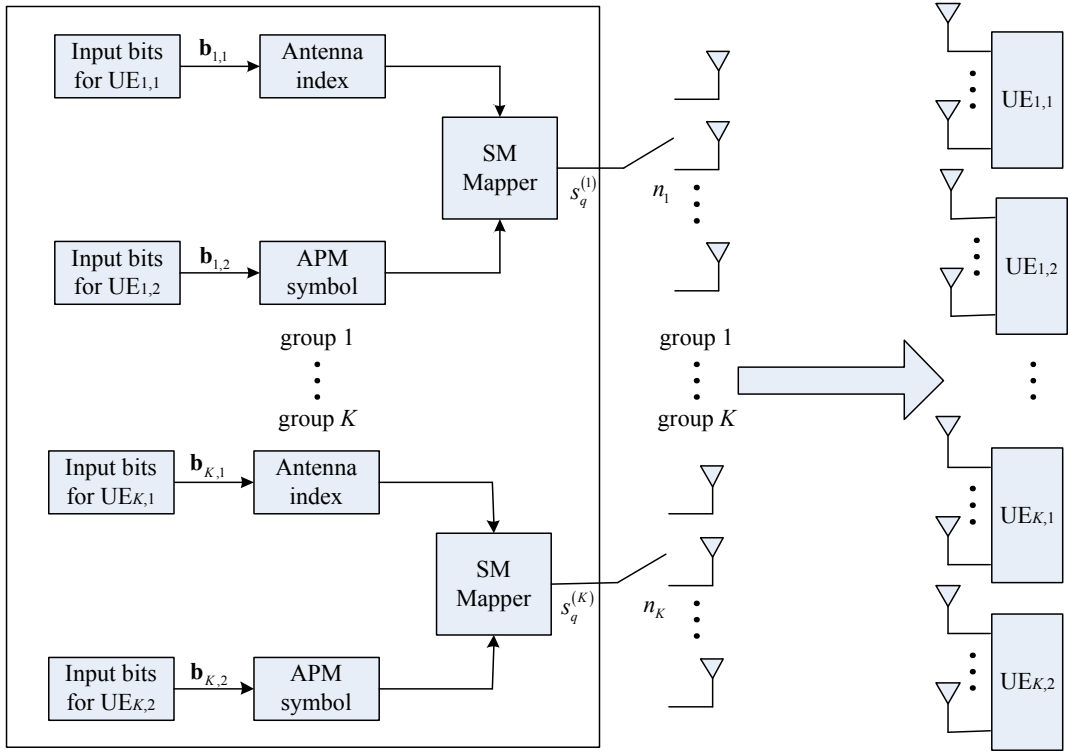}
%  %\hspace{1in}
%  \subfigure[]{\label{fig:1b}\includegraphics[scale=0.6]{system_model_version222.eps}}
  \caption{System model of multiuser SM assisted multi-antenna NOMA.}
  \label{fig:system model of the spatial modulation assisted MIMO NOMA}
\end{figure}

\begin{figure}[!ht]
  \centering
  \includegraphics[scale=0.75]{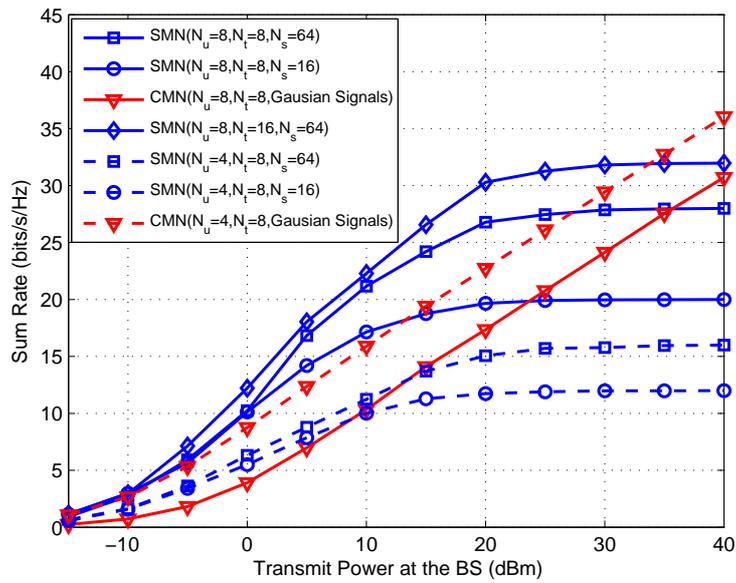}
%  %\hspace{1in}
%  \subfigure[]{\label{fig:1b}\includegraphics[scale=0.6]{system_model_version222.eps}}
  \caption{Sum rate comparison between SM assisted multi-antenna NOMA (SMN) and conventional multi-antenna NOMA (CMN).}
  \label{fig:simulation of sumrate}
\end{figure}

\begin{figure}[!ht]
  \centering
  \includegraphics[scale=0.75]{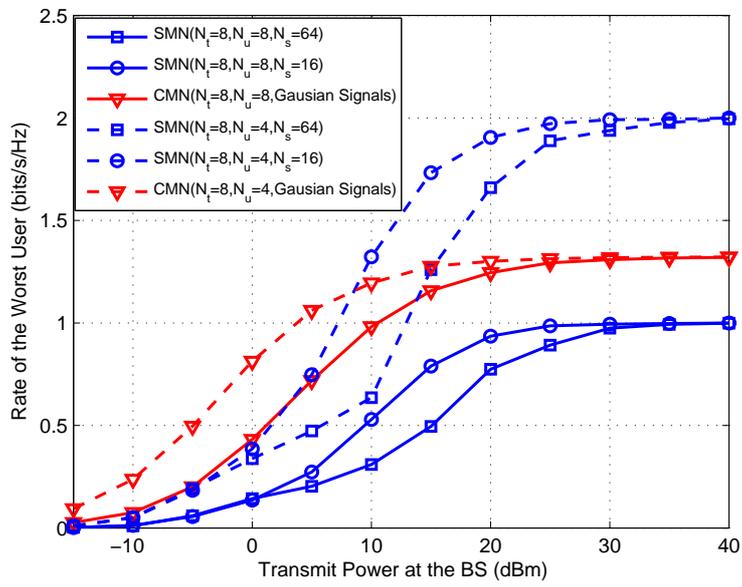}
%  %\hspace{1in}
%  \subfigure[]{\label{fig:1b}\includegraphics[scale=0.6]{system_model_version222.eps}}
  \caption{Worst rate comparison between SMN and CMN.}
  \label{fig:simulation of worst rate}
\end{figure}

\end{footnotesize}
\end{document}